\documentclass{PoS}
\usepackage{amsfonts}
\usepackage{mathrsfs}
\usepackage[ansinew]{inputenc}      
\usepackage{amssymb}                
\usepackage{amsmath}                
\usepackage[]{graphicx}
\usepackage{mathrsfs}
\usepackage{verbatim}
\usepackage{overpic}
\usepackage{xspace}
\usepackage{lineno}
\usepackage{enumitem}
\usepackage[font=small,skip=3mm]{caption}
\usepackage{setspace}
\usepackage{array}
\usepackage{lscape}
\usepackage{rotating}

\newcommand{\ttbar}{\ensuremath{t\bar{t}}\xspace}
\newcommand{\etmiss}{\ensuremath{E \kern-0.6em\slash_{\rm T}}\xspace}
\newcommand{\etmissx}{\ensuremath{E \kern-0.6em\slash_{\rm x}}\xspace}
\newcommand{\etmissy}{\ensuremath{E \kern-0.6em\slash_{\rm y}}\xspace}

\newcommand{\ljets}{\ensuremath{\ell\!+\!{\rm jets}}\xspace}
\newcommand{\dilep}{\ensuremath{\ell\ell}\xspace}

\newcommand{\etal}{\textit{et~al.}\xspace}

\newcommand{\GeV}{\ensuremath{\textnormal{GeV}}\xspace}
\newcommand{\TeV}{\ensuremath{\textnormal{TeV}}\xspace}

\newcommand{\met}{\ensuremath{E_\mathrm{T}^\mathrm{miss}}\xspace}

\newcommand{\fb}{\ensuremath{{\rm fb}^{-1}}\xspace}
\newcommand{\mw}{\ensuremath{M_W}\xspace}
\newcommand{\mt}{\ensuremath{m_t}\xspace}

\newcommand{\kjes}{\ensuremath{k_{\rm JES}}\xspace}

\newcommand{\pt}{\ensuremath{p_{\rm T}}\xspace}

\newcommand{\stwo}{\ensuremath{\sqrt s=1.96~\TeV}\xspace}

\newcommand{\stat}{\ensuremath{{\rm(stat)}}\xspace}
\newcommand{\statjes}{\ensuremath{{\rm(stat\!+\!JES)}}\xspace}
\newcommand{\syst}{\ensuremath{{\rm(syst)}}\xspace}

\title{Direct measurement of the top quark mass in $p\bar p$ collisions at D0}

\ShortTitle{Direct measurement of the top quark mass in $p\bar p$ collisions at D0}

\author{\speaker{Oleg Brandt}\\
       {\bf{\sffamily on behalf of the D0 Collaboration}}\\
        Universit\"at Heidelberg, Kirchhoff-Institut f\"ur Physik, INF 227,\\
        69120 Heidelberg, Germany\\
        E-mail: \email{oleg.brandt@kip.uni-heidelberg.de}}

\abstract{
The mass of the top quark is a fundamental parameter of the Standard Model and has to be determined experimentally. In these proceedings, I review recent direct measurements of the top quark mass in $p\bar p$ collisions at $\sqrt s=1.96$~TeV recorded by the D0 experiment at the Tevatron. The measurements are performed in final states containing one and two charged leptons. I will present the legacy combination of all top quark mass measurements from the D0 experiment and their combination with results from the CDF experiment. A relative precision of down to 0.3\% is attained.
}

\FullConference{The European Physical Society Conference on High Energy Physics\\
         5-12 July\\
         Venice, Italy}

\begin{document}

\section{Introduction}
Since its discovery~\cite{bib:discoverydzero,bib:discoverycdf}, the determination of the top quark mass \mt, a fundamental parameter of the Standard Model (SM), has been one of the main goals of the CERN Large Hadron Collider (LHC) and of the Fermilab Tevatron Collider. Indeed, \mt and masses of $W$ and Higgs bosons are related through radiative corrections that provide a consistency check of the SM~\cite{bib:lepewwg}. Furthermore, \mt dominantly affects the stability of the SM Higgs potential~\cite{bib:vstab1}.
With $\mt=173.34\pm0.76~\GeV$, a world-average combined precision of 0.44\% has been achieved~\cite{bib:combiworld}.

In the SM, the top quark decays to a $W$~boson and a $b$~quark nearly 100\% of the time.
Thus, $\ttbar$ events are classified according to $W$ boson decays as ``dileptonic''~(\dilep), ``lepton+jets'' (\ljets), or ``all--jets''. In the following, I will present recent measurements in the former two channels; a full listing of \mt results from the D0 can be accessed through Ref.~\cite{bib:topresdzero}.

\section{Dilepton channel} \label{sec:ll}

\begin{figure}
\centering
\begin{overpic}[clip,height=4.5cm]{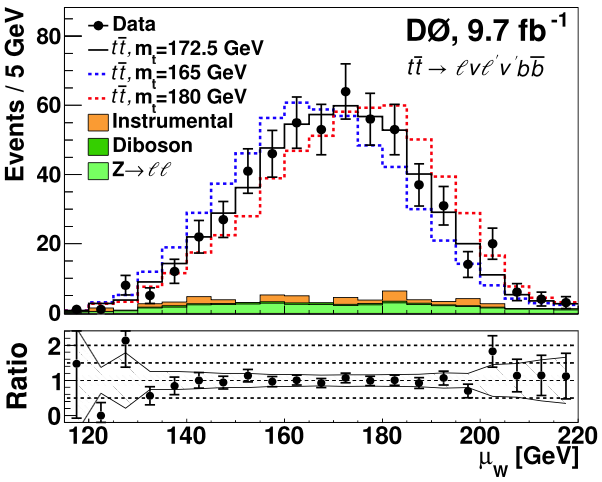}
\put(80,57){\bf\sffamily{(a)}}
\end{overpic}
\qquad
\begin{overpic}[clip,height=4.5cm]{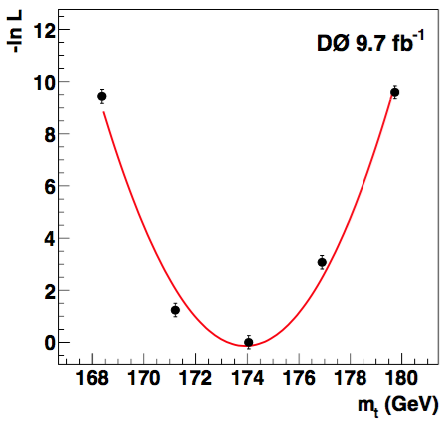}
\put(20,81){\bf\sffamily{(b)}}
\end{overpic}
\caption{
\label{fig:ll}
{\bf(a)} The distribution in the mass estimator $\mu_w$ compared to MC simulations for different \mt hypotheses in the \dilep channel using the neutrino weighting analysis~\cite{bib:d0_ll_nuwt}. 
{\bf(b)} The likelihood in \mt in the \dilep channel using the ME analysis~\cite{bib:d0_ll_me}.
}
\end{figure}

The most precise single measurement of \mt in the \dilep channel at the Tevatron is performed by the D0 Collaboration using 9.7~\fb of $p\bar p$ collisions at \stwo~\cite{bib:d0_ll_nuwt}. The selection requires two isolated leptons ($e$ or $\mu$) of opposite charge, missing transverse momentum \met due to neutrinos, $\geq 2$ jets, where at least one of which is identified as originating from a $b$ quark ($b$-tagged), and other topological selection. Leaving \mt as a free parameter, \dilep final states are kinematically underconstrained by two degrees of freedom. In this analysis, distributions in rapidities of the neutrino and the antineutrino are postulated, and a weight is calculated, which depends on the consistency of the reconstructed $\vec\pt^{\,\nu\bar\nu}\equiv\vec\pt^{\,\nu}+\vec\pt^{\,\bar\nu}$ with the measured missing transverse momentum $\met$ vector, as a function of \mt. D0 uses the first and second moment of this weight distribution to define templates and extract \mt, as shown in Fig.~\ref{fig:ll}~(a) for the first moment $\mu_w$. To reduce the systematic uncertainty, the {\em in situ} jet energy scale (JES) calibration in the \ljets\ channel derived in Ref.~\cite{bib:d0_lj} is applied, accounting for differences in jet multiplicity, luminosity, and detector ageing. The final result reads $m_t=173.32\pm1.36\stat\pm0.85\syst~\GeV$. 
The dominant systematic uncertainties come from the knowledge of the absolute JES (0.47~GeV) and its flavour-dependence (0.36~GeV), and higher order effects on the signal modelling (0.33~GeV).

The top quark mass is also extracted using the matrix element (ME) technique using the same data and a similar selection in the \dilep channel~\cite{bib:d0_ll_me}. This technique determines the probability of observing a given event under both the $\ttbar$ signal and background hypotheses, as a function of \mt~\cite{bib:run1nature}. This probability is calculated {\em ab initio} using the respective MEs of the \ttbar signal and dominant $Z$+jets background, taking into account effects from parton showering (PS), hadronisation, and finite detector resolution. In this analysis, a SM prior is assumed for the transverse momentum distribution of the \ttbar\ system, and the neutrino momenta are integrated over to overcome the challenge of the kinematically underconstrained system. After constraining the JES using \ljets channel results~\cite{bib:d0_lj}, $m_t=173.93\pm1.61\stat\pm0.88\syst~\GeV$ is obtained, as shown in Fig.~\ref{fig:ll}~(b). The systematic uncertainty is dominated by the knowledge of the absolute JES (0.46~GeV) and its flavour-dependence (0.30~GeV), $b$ quark jet identification ($b$-tagging, 0.28~GeV), and hadronisation modelling (0.32~GeV).

\section{Lepton+jets channel} \label{sec:lj}

\begin{figure}[b]
\centering
\begin{overpic}[clip,height=5cm]{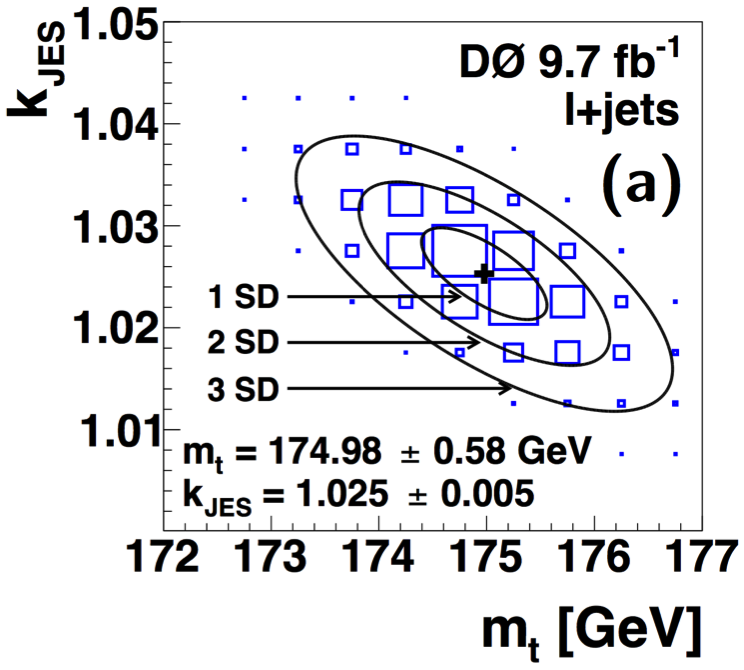}
\end{overpic}
\qquad
\begin{overpic}[clip,height=5cm]{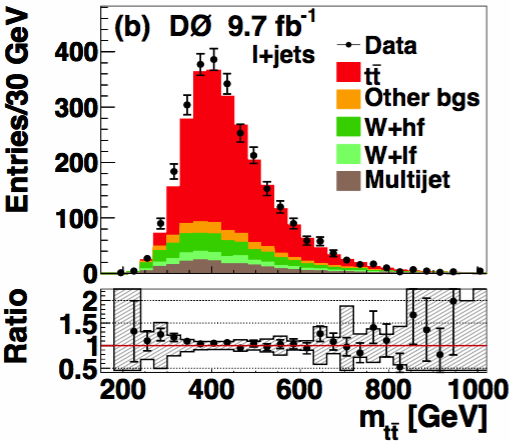}
\end{overpic}
\caption{
\label{fig:lj}
{\bf(a)} The likelihood in $(\mt,\kjes)$ in the \ljets channel using the ME technique~\cite{bib:d0_lj}. Fitted contours of equal probability are overlaid as solid lines. The maximum is marked with a cross. 
{\bf(b)} Invariant mass of the \ttbar system for best-fit values of \mt and \kjes. In the ratio of data to SM prediction, the total systematic uncertainty is shown as a shaded band.
}
\end{figure}

The most precise single measurement of \mt from the Tevatron is performed by the D0 Collaboration using 9.7~\fb of data in the \ljets channel~\cite{bib:d0_lj} with a ME technique. 
The analysis was performed blinded in \mt. This selection requires the presence of one isolated lepton, \met, and exactly four jets with at least one $b$-tag. A new JES calibration from exclusive $\gamma+$jet, $Z+$jet, and dijet events is applied to account for differences in detector response to jets originating from a gluon, a $b$~quark, and $u,d,s,$ or $c$~quarks. 
The overall JES \kjes is calibrated {\it in situ} by constraining the reconstructed invariant mass of the hadronically decaying $W$ boson to $\mw=80.4$~GeV. The likelihood over all candidate events is maximised in $(\mt,\kjes)$ as shown in Fig.~\ref{fig:lj}~(a), and $\mt=174.98\pm0.58\statjes\pm0.49\syst~\GeV$ is obtained.

The estimation of systematic uncertainties is refined through an updated detector calibration, in particular improvements to the $b$-quark JES corrections~\cite{bib:jes}, and using recent improvements in  modeling the \ttbar signal. Furthermore, the statistical component from limited number of MC events is eliminated by reducing the computation time of the ME technique by a factor of 90~\cite{bib:menim}, which results in a more precise estimation of systematic uncertainties. The dominant systematic uncertainties come the modeling of higher-order corrections~(0.15 GeV), hadronisation~(0.26~GeV), and the knowledge of the JES dependence on its flavour~(0.16~GeV) as well as kinematic properties of the jet not captured by \kjes~(0.21~GeV).

\section{Combined results} \label{sec:combo}

\begin{figure}
\centering
\begin{overpic}[clip,height=10cm]{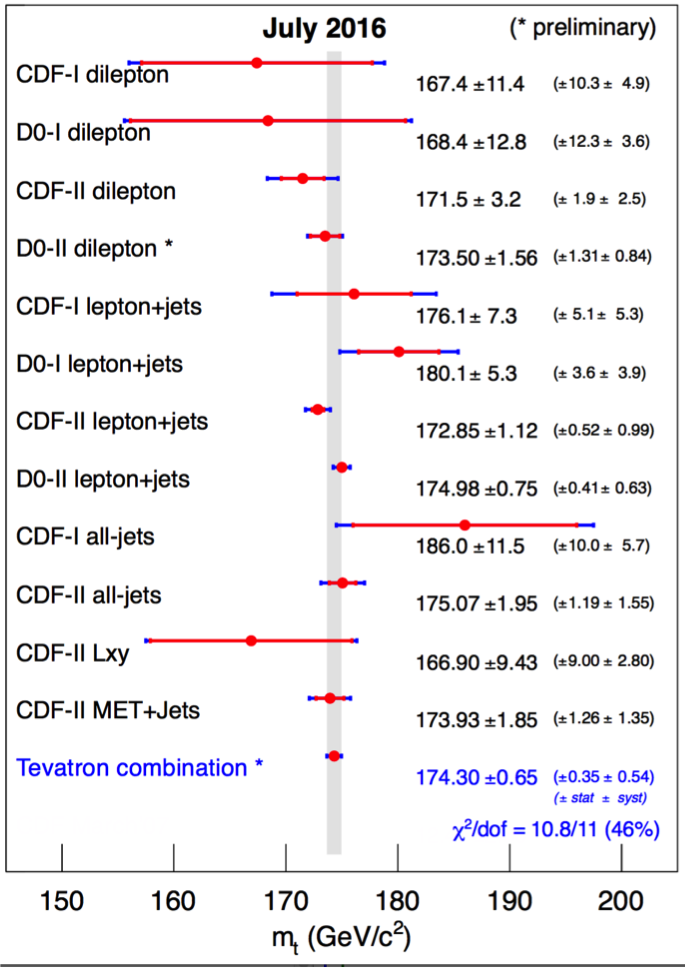}
\end{overpic}
\caption{
\label{fig:overview}
Overview of recent \mt measurements at the Tevatron~\cite{bib:tevatron}. }
\end{figure}

Measurements of \mt from D0 in a given channel in Runs I and II of the Tevatron are statistically combined, accounting for correlations between sources of systematic uncertainty. The final result reads $\mt=174.95\pm0.40\stat\pm0.63\syst~\GeV$~\cite{bib:combo_D0} with a $p$-value of 47\%. A similar combination including \mt results from the CDF Collaboration in Runs I and II, as shown in Fig.~\ref{fig:overview}, results in $\mt=174.30\pm0.35\stat\pm0.54\syst~\GeV$~\cite{bib:tevatron}, corresponding to a relative precision of 0.37\% and a $p$-value of 46\%.

\section{Conclusions}
I presented recent measurements of \mt, a fundamental parameter of the SM. The most precise single measurement at the Tevatron of $\mt=174.98\pm0.58\statjes\pm0.49\syst~\GeV$ is performed by the D0 Collaboration in the \ljets channel, corresponding to a relative precision of 0.43\%. The combination with all other measurements from the D0 experiment results in $\mt=174.95\pm0.40\stat\pm0.63\syst~\GeV$. The Tevatron combination yields $\mt=174.30\pm0.35\stat\pm0.54\syst~\GeV$, which corresponds to a relative precision of 0.37\%.

\section*{Acknowledgments}
I would like to thank my colleagues from the D0 experiment for their help in preparing this article, the staff at Fermilab together with the collaborating institutions, as well as the relevant funding agencies.


\end{document}